%% file: main.tex
\documentclass[preprint,12pt,authoryear]{elsarticle}

\journal{Information and Computation}

\usepackage{amsmath,amssymb,amsthm,amsfonts,mathrsfs,latexsym}
\usepackage{graphicx}
\usepackage{caption}
\usepackage{color}
\usepackage{framed}

\newtheorem{theorem}{Theorem}
\newtheorem{lemma}{Lemma}

\input{macros}

\begin{document}

\begin{frontmatter}



\title{Probabilistic Automata of Bounded Ambiguity}


\author[1,2]{Nathana{\"e}l Fijalkow}
\author[3,5]{Cristian Riveros}
\author[4]{James Worrell}
\address[1]{CNRS, LaBRI}
\address[2]{The Alan Turing Institute of data science, London}
\address[3]{Pontificia Universidad Cat{\'o}lica de Chile}
\address[4]{University of Oxford}
\address[5]{Millennium Institute for Foundational Research on Data (IMFD), Chile}

\begin{abstract}
Probabilistic automata are an extension of nondeterministic finite automata in which transitions are annotated with probabilities.
Despite its simplicity, this model is very expressive and many of the associated algorithmic questions are undecidable.  
In this work we focus on the emptiness problem (and its variant the value problem), which asks whether a given probabilistic automaton accepts some word with probability greater than a given threshold.  
We consider a natural and well-studied structural restriction on automata, namely the degree of ambiguity, which is defined as the maximum number of accepting runs over all words.  
The known undecidability proofs exploits infinite ambiguity and so we focus on the case of finitely ambiguous probabilistic automata.

Our main contributions are to construct efficient algorithms for analysing finitely ambiguous probabilistic automata through a reduction to a multi-objective optimisation problem called the stochastic path problem.
We obtain a polynomial time algorithm for approximating the value of probabilistic automata of fixed ambiguity and a quasi-polynomial time algorithm for the emptiness problem for $2$-ambiguous probabilistic automata.
  
We complement these positive results by an inapproximability result stating that the value of finitely ambiguous probabilistic automata cannot be approximated unless $\PTIME = \NP$.
\end{abstract}


\end{frontmatter}

\section{Introduction}
\input{introduction}

\section{Preliminaries}
\label{sec:preliminaries}

\input{preliminaries}

\section{Undecidability for Linearly Ambiguous Probabilistic Automata}
\label{sec:undecidability}

\input{undecidability}

\section{Hardness of Approximation for Finitely Ambiguous Probabilistic Automata}
\label{sec:hardness}

\input{hardness}

\section{Decidability and Complexity of Finitely Ambiguous Probabilistic Automata}
\label{sec:properties}

\input{properties}

\section{Algorithms and Approximations for Finitely Ambiguous Probabilistic Automata}
\label{sec:emptiness}

\input{algorithms}

\section*{Acknowledgements}
We thank Shaull Almagor for reporting a mistake in the proof (and statement) of Theorem 5.
We also thank Itay Hasson and his advisor for fixing a typo in the same proof.

\section*{Conclusions}

We have initiated the study of the computational complexity of analysing finitely ambiguous probabilistic automata.
Our main conceptual tool is a reduction to a multi-objective optimisation problem called the stochastic path problem.
There remain many gaps in complexity, leaving interesting open problems.
The most exciting is the complexity of the emptiness problem for $k$-ambiguous probabilistic automata:
can it be solved in polynomial time for $k = 2$, or in quasi-polynomial time for every $k > 2$?


\bibliographystyle{elsarticle-harv} 
\bibliography{biblio}

\end{document}

%% file: macros.tex
\renewcommand{\epsilon}{\varepsilon}
\renewcommand{\phi}{\varphi}

\newcommand{\set}[1]{\left\{ #1 \right\}}

\newcommand{\Q}{\mathbb{Q}}
\newcommand{\N}{\mathbb{N}}

\newcommand{\NEXPTIME}{\mathbf{NEXPTIME}}
\newcommand{\PSPACE}{\mathbf{PSPACE}}
\newcommand{\NP}{\mathbf{NP}}
\newcommand{\NC}{\mathbf{NC}}
\newcommand{\PTIME}{\mathbf{P}}

\newcommand{\C}{\mathcal{C}}

\newcommand{\Output}{\text{Output}}

\renewcommand{\P}{\mathcal{P}}

\newcommand{\D}{\mathcal{D}}
\newcommand{\Run}{\operatorname{Run}}
\newcommand{\val}{\text{val}}
\newcommand{\bin}{\mathrm{bin}}

\newcommand{\MaxClique}{\mathrm{MaxClique}}

%% file: introduction.tex
\textit{Probabilistic automata} are a natural extension of
non-deterministic automata that were introduced by~\cite{Rabin63}.
Such automata 
can also be seen as  a type of weighted automata, as defined by~\cite{Schutzenberger61},
over the semiring of real numbers.
Syntactically, a probabilistic automaton is a
non-deterministic finite automaton in which each edge is annotated by
a probability.  Such an automaton associates to every word a value
between $0$ and $1$, which is the total probability that a run on the
word ends in an accepting state.  We call this the acceptance
probability of the word.

Despite their simplicity, probabilistic automata are very expressive
and have been widely studied.  Unfortunately the price of this
expressiveness is that almost all natural decision problems
are undecidable.  Consequently, various 
approaches based on subclasses of probabilistic automata determined by bounds on resources, such as structure, dimension, or randomness, have
been studied~\cite{CT12,FGO12,FGKO15,CSV17}.

In this paper, we look at probabilistic automata of \textit{bounded ambiguity}, 
where the ambiguity of a word relative to a given automaton is the number of accepting
runs.  We say that a probabilistic automaton is $f$-ambiguous, for a
function $f : \N \to \N$, if every word of length $n$ has at most $f(n)$ accepting runs.  
(Note that ambiguity is a property of the underlying nondeterministic finite automata, and is independent of the transition probabilities.)  
This notion has been extensively studied in automata theory; in particular, the landmark paper of~\cite{WS91} 
gives respective structural characterisations of
the respective classes of finitely, polynomially, and exponentially ambiguous
nondeterministic finite automata, from which polynomial-time
algorithms are obtained for deciding membership in each of these
classes.

We focus on the most natural and well-studied problem for
probabilistic automata, called the \textit{emptiness problem}: given a
probabilistic automaton and a threshold, does there exist a word
accepted with probability exceeding a given threshold?  
Since the emptiness problem is already undecidable for linearly ambiguous 
probabilistic automata (as shown by~\cite{DJLMPW18,CSV18}), 
we focus on finitely ambiguous probabilistic automata.

We study the complexity of the emptiness problem on various classes of
finitely ambiguous probabilistic automata.  For each positive integer
$k$ we consider the class of \emph{$k$-ambiguous probabilistic automata},
\textit{i.e.}, automata with at most $k$ accepting runs on any word.  
More generally we fix a polynomial $p$ and consider the class of automata
whose ambiguity is at most $p(n)$, where $n$ is the number of states.
More generally still, bearing in mind that the ambiguity can be
exponential in the number of states, we have the class of all finitely
ambiguous automata.

Our main results are as follows.  We show that the emptiness problem
for finitely ambiguous probabilistic automata is, respectively:
\begin{itemize}
	\item in $\NEXPTIME$ and $\PSPACE$-hard for the class of all finitely ambiguous automata;
	\item $\PSPACE$-complete for the class of probabilistic automata with
          ambiguity bounded by a fixed non-constant polynomial in the
          number of states.
	\item  in $\NP$ for the class of $k$-ambiguous probabilistic automata, for every positive integer $k$.
	\item in quasi-polynomial time for the class of $2$-ambiguous
          probabilistic automata.
\end{itemize}

A natural counterpart of the emptiness problem is the function
problem of computing the value of a probabilistic automaton, that is,
the supremum over all words of the acceptance probability of a word.
Here we show:
\begin{itemize}
\item for the class of all finitely ambiguous probabilistic automata, there is
  no polynomial-time approximation algorithm for the value problem
unless $\PTIME = \NP$,
\item for each fixed $k$, the value of a $k$-ambiguous probabilistic automaton
  is approximable up to any multiplicative constant in polynomial
  time.
\end{itemize}

The starting point to prove these results is to give an upper bound on
the length of a witness word, \textit{i.e.} whose probability exceeds a given
threshold.  More precisely, we show that for a $k$-ambiguous
probabilistic automaton with $n$ states there is a word of length at most $n^k$
reaching the maximal probability. 
More generally, we show that for a finitely ambiguous probabilistic automaton with $n$ states, 
there is such a word of length at most $n!$. 
Both results lead to complexity upper bounds for the emptiness problem of finitely ambiguous probabilistic automata
in various regimes of ambiguity.
Most of the remainder of the paper is devoted to the case of $k$-ambiguous automata for a fixed $k$.

We give a polynomial-time reduction from the emptiness problem for
$k$-ambiguous probabilistic automata to a multi-objective optimisation
problem, which we call the \emph{$k$-stochastic path problem}.  Using
this reduction, we obtain a polynomial-time algorithm for
approximating the value of a $k$-ambiguous probabilistic automata, and
a quasi-polynomial time algorithm for the emptiness problem of
$2$-ambiguous probabilistic automata.

%% file: preliminaries.tex
Given $a,b \in \N$, we write $[a,b]$ for $\set{a,a+1,\dots,b}$.

Let $\Sigma$ be a finite alphabet. For any word $w \in \Sigma^*$, we let $|w|$ denote its length.  
Given a finite set $Q$, a (sub)distribution is a function $\delta : Q \rightarrow [0,1]$ such that
$\sum_{q \in Q} \delta(q) \le 1$.  
We let $\D(Q)$ denote the set of distributions over $Q$.

A \emph{probabilistic automaton} is a tuple
$\P = (Q, q_{in}, \Delta, F)$, where $Q$ is a finite set of states,
$q_{in}$ is the initial state, $\Delta : Q \times \Sigma \to \D(Q)$ is the
transition function, and $F$ is the set of accepting states.  Given a
word $w = a_1 \cdots a_n$, a run $\rho$ over $w$ is a sequence of
states $q_0, q_1, \ldots, q_n$.  The probability of such a run is
$\P(\rho) = \prod_{\ell \in \set{1,\ldots,n}} \Delta(q_{\ell - 1},a_\ell)(q_\ell)$.  
We let $\Run_\P(p \xrightarrow{w} q)$ denote the set of runs $\rho$ over $w$ starting in $p$ and finishing in $q$ with $\P(\rho) > 0$.  
The number $\P(p \xrightarrow{w} q)$ is the probability to go from $p$ to $q$ reading $w$, defined as the sum of
the probabilities of its runs, namely:
\[
\P(p \xrightarrow{w} q) = \sum_{\rho \in \Run_\P(p \xrightarrow{w} q)} \P(\rho).
\]
A run $\rho$ is accepting if it starts in $q_{in}$, satisfies $\P(\rho) > 0$, and finishes in an accepting state, \textit{i.e.} a state in $F$.  
We let $\Run_\P(w)$ denote the set of accepting runs over $w$. 
The \textit{probability} of $w$ over $\P$ is defined as the sum of the probabilities of its accepting runs by:
\[
\P(w) = \sum_{\rho \in \Run_\P(w)} \P(\rho).
\]

It is sometimes convenient to have instead of one initial state a distribution of initial states.
The definitions above are easily extended to this setting.
This extension comes at no cost as we can transform such a probabilistic automaton 
into one with a single initial state by adding one state and performing some normalisation 
(specifically, removing an $\varepsilon$-transition).

\medskip\noindent\textbf{Ambiguity.}  In this paper, we consider
different subclasses of probabilistic automata, obtained by
restrictions on \emph{ambiguity}.  More specifically, we say that:
\begin{itemize}
	\item $\P$ is \emph{unambiguous} if every word $w$ has at most one accepting run, \textit{i.e.} $|\Run_\P(w)| \le 1$.
	
	\item $\P$ is $k$-\emph{ambiguous} if every word $w$ has at most $k$ accepting runs, \textit{i.e.} $|\Run_\P(w)| \le k$.

	\item $\P$ is \emph{finitely ambiguous}, if there exists $k$ such that $\P$ is $k$-ambiguous.
	
	\item $\P$ is \emph{polynomially ambiguous}, if there exists a polynomial $P$ such that for every word $w$, we have $|\Run_\P(w)| \le P(|w|)$.
\end{itemize}
If the polynomial $P$ is linear or quadratic then we say that a polynomially ambiguous automaton $\P$ is linearly ambiguous or quadratically ambiguous, respectively.
It is proved in~\cite{WS91} that it is decidable in polynomial time whether an automaton $\P$ is unambiguous, finitely ambiguous, 
or polynomially ambiguous.
Furthermore, a consequence of the results of~\cite{WS91} is that an automaton which is not finitely ambiguous has ambiguity
bounded below by a linear function.

\medskip\noindent\textbf{Emptiness problem and value.}
Let $\P$ be a probabilistic automaton and $c$ a threshold. 
Following~\cite{Rabin63},
we define the threshold language induced by $\P$ and $c$ as:
\[
L^{> c}(\P) = \set{w \in \Sigma^* \mid \P(w) > c}.
\]
The \emph{emptiness problem} asks, given a probabilistic automaton
$\P$ and a threshold $c$, whether the language $L^{> c}(\P)$ is
non-empty, that is, whether there exists a word $w$ such that
$\P(w) > c$.

A related function problem is to compute the value of a probabilistic automaton $\P$, defined by
$\val(\P) = \sup_{w \in \Sigma^*} \P(w)$.
Note that the emptiness problem is equivalent to asking whether $\val(\P) > c$.

%% file: undecidability.tex
In this section, we discuss undecidability results for linearly ambiguous probabilistic automata, 
which justify the focus of our paper on finitely ambiguous probabilistic automata.

\begin{theorem}[\cite{CSV18,DJLMPW18}]
\label{thm:undecidable}
The emptiness problem is undecidable for linearly ambiguous
probabilistic automata.
\end{theorem}

Undecidability of the emptiness problem has long been known for
general probabilistic automata, see~\cite{Paz71,Bertoni74,GO10}. 
However, the automata involved in the proof have exponential ambiguity.

In the conference version of this paper we explained how to adapt the proof strategy above to obtain
the undecidability of the emptiness problem for quadratically ambiguous probabilistic automata, see~\cite{FRW17}.
We left open whether the undecidability already holds for linearly ambiguous automata.
Two subsequent independent papers filled the gap by showing the stronger result stated in Theorem~\ref{thm:undecidable},
see~\cite{DJLMPW18,CSV18}.
We therefore focus here on the isolation problem.

Given a probabilistic automaton $\P$, we say that a threshold $c$ is
isolated if there exists $\varepsilon > 0$ such that for all words
$w$, we have $|\P(w) - c| > \varepsilon$. 
\cite{Rabin63} proved that if a threshold $c$ is isolated then the corresponding
language $L^{> c}(\P)$ is regular.  The isolation problem asks to
determine whether a given threshold is isolated for a given automaton.
This problem was shown to be undecidable by~\cite{Bertoni74};
we refer to~\cite{Fijalkow17} for a new presentation of this result.
We can refine the result of~\cite{Bertoni74} to obtain:

\begin{theorem}
  The isolation problem is undecidable for linearly ambiguous
  probabilistic automata.
\end{theorem}

We start by describing the key ingredient in the undecidability proof of~\cite{Bertoni74},
which is the construction of a probabilistic automaton computing the value of a rational number given
in binary with least significant digit on the left:
\[
\bin^R(a_1 \cdots a_n) = \sum_{i = 1}^n \frac{a_i}{2^{n-i+1}}.
\]
The automaton proposed by Bertoni has exponential ambiguity. 
However, it is possible to construct a linearly ambiguous probabilistic automaton computing the same function but \textit{reversing} the input:
\[
\bin(a_1 \cdots a_n) = \sum_{i = 1}^n \frac{a_i}{2^i}.
\]
\begin{figure}
\centering
\includegraphics[scale=.8]{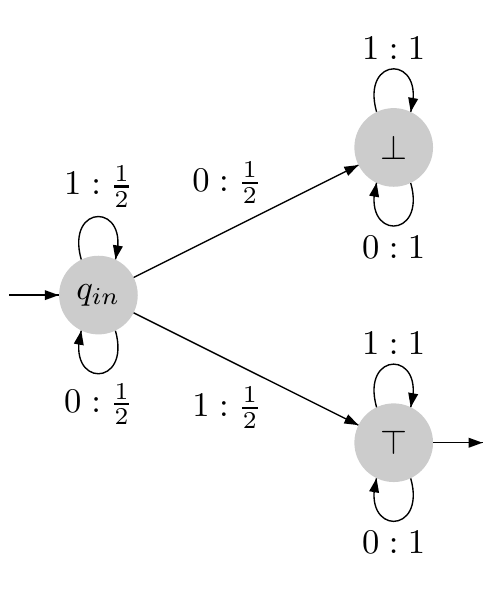}
\caption{\label{fig:binary_pa}A probabilistic automaton computing $\bin$.}
\end{figure}
The automaton is represented in Figure~\ref{fig:binary_pa}.

\begin{proof}
  We construct a reduction from a variant of the Post's Correspondence Problem (PCP), called the infinite PCP, and shown to be undecidable
  by~\cite{Ruohonen85}.  The problem asks, given two homomorphisms
  $\varphi_1, \varphi_2 : \Sigma^* \rightarrow \set{0,1}^*$, to decide whether
  there exists an infinite word $w$ in $\Sigma^\omega$ such that
  $\varphi_1(w) = \varphi_2(w)$ (where $\varphi_1,\varphi_2$ are
  extended to maps on $\Sigma^\omega$).
  We first observe that equivalently, we ask
  whether for every $\varepsilon > 0$ there exists a non-empty finite word $w$ such that
  $|\bin(\varphi_1(w)) - \bin(\varphi_2(w))| \le \varepsilon$.

Indeed, if there exists an infinite word $w$ such that $\varphi_1(w) = \varphi_2(w)$, 
then the sequences obtained by considering the images under $\varphi_1$ and $\varphi_2$ of prefixes of $w$ have arbitrarily long common prefixes, 
so the difference of their binary values converges to $0$.
Conversely, assume that for any $\varepsilon > 0$ there exists a non-empty finite word $w$ such that 
$|\bin(\varphi_1(w)) - \bin(\varphi_2(w))| \le \varepsilon$, 
then we construct a solution to the infinite PCP using K\"{o}nig's lemma.
To this end, for each $n$ let $w_n$ be a finite word such that $|\bin(\varphi_1(w_n)) - \bin(\varphi_2(w_n))| < 2^{-n}$, 
\textit{i.e.}, such that $\varphi_1(w_n)$ and $\varphi_2(w_n)$ coincide on the first $n$ letters.
Applying K\"{o}nig's Lemma to the infinite tree defined by the prefix closure of the set $\set{w_n \mid n \geq 0}$ 
(\textit{i.e.}, each node in the tree is the prefix of some word $w_n$), there exists an infinite word $w$ such that $\varphi_1(w) = \varphi_2(w)$.

We now construct a reduction from the infinite PCP to the isolation problem for linearly ambiguous probabilistic automata.
Let $\varphi_1$ and $\varphi_2$ be two homomorphisms,
we construct a linearly ambiguous probabilistic automaton $\P$ such that for every non-empty word $w$,
\[
\P(w) = \frac{1}{2} \left( \bin(\varphi_1(w)) + 1 - \bin(\varphi_2(w)) \right).
\]
Let us fix $\varepsilon > 0$ and $w$ a non-empty word.
The following equivalence holds
\[
|\bin(\varphi_1(w)) - \bin(\varphi_2(w))| \le \varepsilon \Longleftrightarrow |\P(w) - \frac{1}{2}| \le \varepsilon,
\]
implying that the answer to the infinite PCP problem of $(\varphi_1,\varphi_2)$ is positive
if and only if the threshold $\frac{1}{2}$ is not isolated for $\P$.

To construct the automaton $\P$, we proceed as follows.
Let~$\P_0$ be the automaton computing $\bin$.
First, we construct~$\P_{\phi_1}$, which is obtained from~$\P_0$ by extending the transition function as follows:
when reading the letter~$a$, the automaton $\P_{\phi_1}$ simulates the transitions of $\phi_1(a)$, which is a word over~$\set{0,1}^*$.
By construction we have $\P_{\phi_1}(w) = \bin(\phi_1(w))$, and $\P_{\phi_1}$ is linearly ambiguous.
We construct $\P_{\phi_2}$ similarly, but complementing the set of accepting states, so that $\P_{\phi_2}(w) = 1 - \bin(\phi_2(w))$.
Finally, $\P$ is the disjoint union of $\P_{\phi_1}$ and $\P_{\phi_2}$,
with a distribution of initial states assigning probability $\frac{1}{2}$ to each of the initial states of $\P_{\phi_1}$ and $\P_{\phi_2}$.
The automaton $\P$ is linearly ambiguous, and as explained in the previous section, 
one can easily transform it into a (still linearly ambiguous) probabilistic automaton with a single initial state.
\end{proof}

An automaton is either finitely ambiguous, or at least linearly ambiguous.
Bearing in mind our undecidability results for linearly ambiguous automata, we are led to focus
on decidability results for finitely ambiguous automata.

%% file: hardness.tex
In this section we show another negative result for finitely ambiguous probabilistic automata.
The question we ask is whether there exists an approximation algorithm for computing the value of such automata, in the following sense:
a $K(n)$-approximation algorithm takes as input a probabilistic automaton $\P$ with $n$ states
and outputs $v$ such that
\[
\frac{\val(\P)}{K(n)} \le v \le K(n) \cdot \val(\P).
\]
In other words, we consider approximation algorithms up to a $K(n)$ multiplicative factor.

The following hardness of approximation result complements the positive results obtained later in this paper, 
witnessing the complexity of analysing even finitely ambiguous probabilistic automata.

\begin{theorem}[\cite{LP02}]\label{thm:hardness_approximation}
	For every $\varepsilon > 0$, there is no polynomial time $O(n^{\frac{1}{2} - \varepsilon})$-approximation algorithm 
	for the value of finitely ambiguous probabilistic automata, unless $\PTIME = \NP$.
\end{theorem}

The proof is a direct adaptation of the reduction constructed by~\cite{LP02}, which uses the similar but different framework of 
Hidden Markov models.

\begin{proof}
We construct a reduction from the clique problem to the value of finitely ambiguous probabilistic automata.
The clique problem asks, given an undirected graph $G$, to compute the size of a largest clique,
\textit{i.e.} a subset of vertices such that there is an edge between any two vertices.

Given a graph $G$ with $n$ vertices, we construct an $n$-ambiguous probabilistic automaton $\P_G$ 
with $O(n^2)$ states such that for each $m$ smaller than $n$, there exists a word $w$ such that $\P_G(w) \ge \frac{m}{n 2^{n-1}}$
if and only if the graph $G$ contains a clique of size at least $m$.

We write $V = \set{v_1,\dots,v_n}$ for the set of vertices of $G$ and $E \subseteq V \times V$ the set of edges of $G$.
By convention $G$ does not contain self-loops, \textit{i.e.} $(v,v) \notin E$.
The set of states of $\P_G$ is $\set{v_{i,j} : i \in [1,n], j \in [0,n]}$.
The transition function is defined as follows for $i,j \in [1,n]$:
\[
\Delta(v_{i,j-1},1)(v_{i,j}) = 
\begin{cases}
1 & \text{if } j = i \\
\frac{1}{2} & \text{if } (i,j) \in E \\
0 & \text{if } (i,j) \notin E \\
\end{cases}
\quad 
\Delta(v_{i,j-1},0)(v_{i,j}) = 
\begin{cases}
\frac{1}{2} & \text{if } (i,j) \in E \\
0 & \text{ if } (i,j) \notin E \\
\end{cases}
\] 
The set of accepting states is $\set{v_{i,n} : i \in [1,n]}$.
The automaton $\P_G$ has a distribution of initial states: $\frac{1}{n}$ for each $v_{i,0}$ with $i \in [1,n]$.

We remark that only words of length exactly $n$ have accepting runs.
Such words are in bijection with subsets of vertices:
the word $w = a_1 \dots a_n$ corresponds to the subset of vertices $S_w = \set{v_i : a_i = 1}$.
The automaton $\P_G$ on input $w$ has $n$ runs, one for each vertex $v_i$, chosen each with probability $\frac{1}{n}$.
Each accepting run has probability $\frac{1}{2^{n-1}}$, because the probability is divided by $2$ at each transition
except for one transition (case $i = j$ in the definition of $\Delta$).
The run over $w$ corresponding to a vertex $v_i$ is accepting if and only if 
$v_i$ is in $S_w$ and all vertices of $S_w$ are neighbours of $v_i$.
Consequently, the set of vertices corresponding to accepting runs form a clique (included in $S_w$).
Hence a clique of size $m$ induces a word accepted with probability $\frac{m}{n 2^{n-1}}$, and conversely.

Let $\MaxClique(G)$ denote the size of a largest clique in $G$, the equivalence above reads $\MaxClique(G) = n 2^{n-1} \val(\P_G)$.
It follows that a $K(n)$-approximation algorithm for the value of finitely ambiguous probabilistic automata
induces a $K(n^2)$-approximation algorithm for the size of a largest clique.
Indeed, given a graph $G$ with $n$ vertices, we construct the probabilistic automaton $\P_G$ (recall that it has $O(n^2)$ states), 
and run the $K(n)$-approximation algorithm, yielding a $K(n^2)$-approximation of the value of $\P_G$, 
which multiplied by $n 2^{n-1}$ yields a $K(n^2)$-approximation of the size of a largest clique.
\cite{Zuckerman07} proved that for every $\varepsilon > 0$,
there is no polynomial time $O(n^{1 - \varepsilon})$-approximation algorithm for the size of a largest clique, unless $\PTIME = \NP$, 
implying our result.
\end{proof}

\begin{figure}
\centering
\includegraphics[scale=.8]{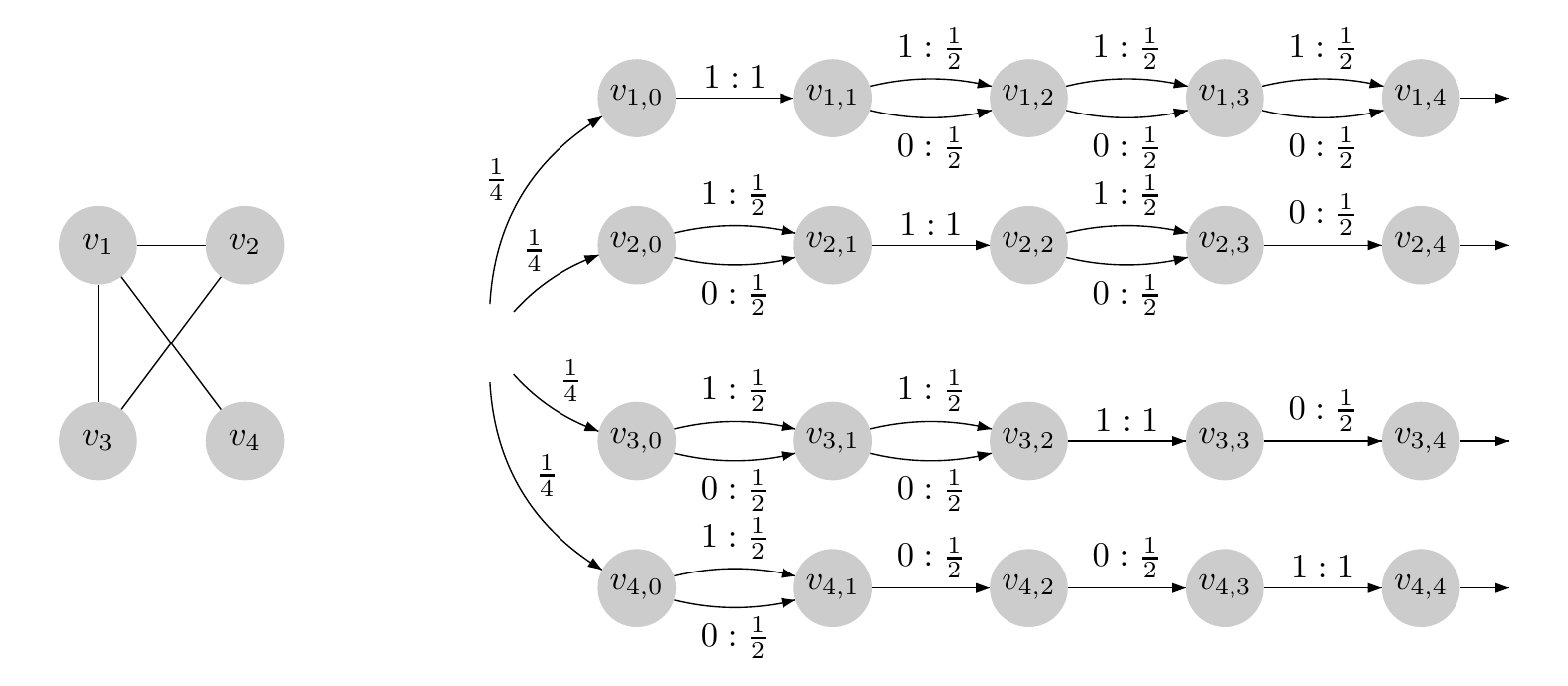}
\caption{\label{fig:reduction}On the left a graph $G$ and on the right the corresponding finitely ambiguous probabilistic automaton $\P_G$
such that $\MaxClique(G) = 4 \cdot 2^{3} \cdot \val(\P_G)$.}
\end{figure}

In Figure~\ref{fig:reduction} we illustrate this construction with a graph $G$ with four vertices $v_1, \ldots, v_4$ and the corresponding finitely ambiguous probabilistic automaton $\P_G$. 
For example, here the word $1110$ represents the set of vertices $\set{v_1,v_2,v_3}$ which has probability $\frac{3}{4 \cdot 2^3}$ in $\P_G$, and is indeed a clique with three vertices.

%% file: properties.tex
In this section we study the emptiness problem for finitely ambiguous probabilistic automata.
We start by showing regularity of the threshold language $L^{> c}(\P)$ for a finitely
ambiguous probabilistic automaton $\P$ and a threshold $c$.  
A classical result due to~\cite{Rabin63} shows that the threshold language need not be regular in general. 
Unfortunately our proof of regularity, while constructive, is not useful for determining the
complexity of the emptiness problem.  However we are able to give
a direct simple argument that bounds the length of witnesses for the
emptiness problem.  We then use these bounds to analyse the complexity
of the emptiness problem.

Our proof of regularity makes use of the theory of well quasi orders.
We refer to~\cite{S17} for a survey on the use of this theory in theoretical computer science.
We are only concerned with the well quasi order over $\N^k$ induced by the pointwise order written $\le$.
A subset $S$ of $\N^k$ is \textit{downward closed} if $x \in S$ and $y \le x$ implies $y \in S$,
and it is \textit{directed} if for any $x,y \in S$ there exists $z \in S$ such that $x \le z$ and $y \le z$.
An \emph{ideal} is a subset of $\N^k$ which is both downward closed and directed.  
Every ideal $I$ has the form
\begin{gather}
I = \set{(n_1,\ldots,n_k) \in \N^k : n_{i_1} \leq a_1 \wedge \ldots \wedge n_{i_s} \leq a_s} 
\label{eq:idealrep}
\end{gather}
for certain indices $1 \leq i_1 < \ldots < i_s \leq k$ and natural numbers $a_1,\ldots,a_s$.  
From the fact that $\N^k$ is a well quasi order it follows that every downward closed subset $D \subseteq \N^k$ 
can be written as a finite union of ideals.
Such a decomposition can be computed from the finite set of minimal elements of $\N^k \setminus D$ as explained in~\cite{lazic2015ideal}.

\begin{theorem}
	Let $\mathcal{P}$ be a finitely ambiguous probabilistic automaton
	and $c$ a threshold.  Then $L^{> c}(\P)$ is a regular language.
\end{theorem}
\begin{proof}
	Let $\P = (Q, q_{in}, \Delta, F)$ be a $k$-ambiguous probabilistic automaton.
	A transition is a triple $(p,a,q) \in Q\times\Sigma\times Q$ such that $\Delta(p,a)(q)>0$. 
	Let $s$ denote the number of transitions of $\P$, we fix a linear ordering on these transitions.
	We say that $m = (m_{i,j}) \in \N^{s \times k'}$ is \emph{admissible} for a word $w\in \Sigma^*$ if there exist $k'$
	(distinct) accepting runs of $\P$ on $w$ such that $m_{i,j}$ is the number of times that the $i$-th transition is taken in the
	$j$-th accepting run.
	
	For any ideal $I \subseteq \N^{s \times k'}$ the following language is regular:
	\[
	\set{w : \exists m \in I \text{ admissible for } w}.
	\]
	A non-deterministic automaton for this language guesses $k'$ accepting runs of $\P$ and counts the number of times each transition is
	taken on each accepting run up to a finite threshold $N$, where $N$ is the largest integer appearing in the description of $I$ in the
	form (\ref{eq:idealrep}).  
	It follows that for any downward closed	subset $D \subseteq \N^{s \times k'}$, the following language is regular:
	\[
	\set{w : \exists m \in D \text{ admissible for } w}.
	\]
	
	Now let $\lambda_1,\ldots,\lambda_s$ be the transition probabilities
	occuring in $\P$, listed according to the ordering on the transitions.
	Given $k' \in \N$, consider the set of tuples
	\[ 
	S_{k'} \ = \ \left\{ 
	(m_{i,j}) \in \N^{s \times k'} \ : \  
	\sum_{j=1}^{k'} \lambda_1^{m_{1,j}} \dots \lambda_s^{m_{s,j}} > c 
	\right\}. 
	\] 
	For any word $w \in \Sigma^*$, we have $w \in L^{> c}(\P)$ if and only if 
	there exists $k' \le k$ and $m \in S_{k'}$ that is admissible for $w$.  
	Since each set $S_{k'}$ is downward closed, it follows that $L^{> c}(\P)$ is regular.
\end{proof}

The threshold language $L^{> c}(\P)$ of a finitely ambiguous probabilistic automaton is regular, however, this does not say anything about how to decide efficiently whether $L^{> c}(\P)$ is empty or not. 
We say that a word $w$ is a non-emptiness witness, or simply a witness, if $w \in L^{> c}(\P)$.
The next step is to bound the length of witnesses whenever $L^{> c}(\P) \neq \emptyset$. 
This will lead to upper bounds on the complexity of the emptiness problem.

\begin{lemma}\label{lem:pumping_emptiness}
	Let $\P$ be a $k$-ambiguous	probabilistic automaton with $n$ states.  
	For every word $w$, there exists a word $w'$ of length at most $n^k$ such that $\P(w) \le \P(w')$.
	This implies that the value of $\P$ is reached by some word of length at most $n^k$.
\end{lemma}

\begin{proof}
	Let $\P = (Q, q_{in}, \Delta, F)$ and suppose that there are exactly $k'$ accepting runs on $w$ for some
	$k'\leq k$.  If $w$ has length strictly greater than $n^{k'}$ then
	there exists a factorization $w=xyz$ for $x,y,z \in \Sigma^*$, with
	$y$ non-empty and $xz$ of length at most $n^k$, such that for each of the accepting runs on $w$, the
	infix corresponding to the factor $y$ starts and ends in the same state.  
	Then we have
\[
\renewcommand{\arraystretch}{1.5}
\begin{array}{lll}
\P(w) & =   & \sum_{q \in F}\sum_{p \in Q} \P(q_{\mathit{in}} \xrightarrow{x} p) \P(p \xrightarrow{y} p) \P(p \xrightarrow{z} q) \\
	  & \le & \sum_{q \in F}\sum_{p \in Q} \P(q_{in} \xrightarrow{x} p)  \P(p \xrightarrow{z} q) \\
	  & =   & \P(xz) \, .
\end{array}
\]
\end{proof}

Note that if $k$ is fixed, then the length of a witness for $L^{> c}(\P)$ is polynomial in the number of states of the automaton.
Unfortunately, it has been shown in~\cite{WS91} that the ambiguity of a finitely ambiguous automaton can be exponential in the number of
states and, thus, the previous lemma gives a double exponential bound for a witness of $L^{> c}(\P)$ when $k$ is not fixed.  
The next result shows that the length of a witness is at most exponential in the number of states.

\begin{theorem}\label{theo:pumping_emptiness}
	Let $\P$ be a finitely ambiguous probabilistic automaton with $n$ states.  For every word $w$, there
	exists a word $w'$ of length at most $(n+1)!$ such that
	$\P(w) \le \P(w')$.
	This implies that the value of $\P$ is reached by some word of length at most $(n+1)!$.
\end{theorem}

 \begin{proof}
   Consider a word $w = a_1 \cdots a_\ell$ of length at least $(n+1)!$.  
   For any $i \in [1,\ell]$, let $R_i$ be the set of states reached when reading the prefix of $w$ of length $i$ and participating in at least one accepting run over $w$.  
   We equip $R_i$ with the order defined by $p \leq q$ if 
   \[
   \P(q_{in} \xrightarrow{a_1 \cdots a_i} p) \le \P(q_{in} \xrightarrow{a_1 \cdots a_i} q),
   \]
   \textit{i.e.}, after reading the prefix $a_1 \cdots a_i$ of $w$ the probability of being in state $p$ is at most that of being in state $q$.
   Assume that ties are resolved consistently for all $R_i$.
   There are $(n+1)!$ possible values for the ordered sets $R_i$: there are $n!$ total orders on $n$ elements, multiplied by $n+1$ possibilities for the first state in $R_i$.

 	Since $w$ has length at least $(n+1)!$, there exist two positions $i < j$ such that the ordered sets $R_i$ and $R_j$ coincide, 
 	let us refer to their common value as $R$. 
 	There exists a factorization $w = x y z$, with $y$ the word in between positions $i$ and $j$.
 	Then we look at the runs of $y$ from $R$ to $R$, and make the following claims:
 	\begin{enumerate}
 		\item For every $p \in R$, there exists a run over $y$ from $p$ to a state in $R$.
 		\item For every $p \in R$, there exists at most one run over $y$ from $p$ to a state in $R$.
 		\item For every $p \in R$, we have $\P(q_{in} \xrightarrow{xy} p) \le \P(q_{in} \xrightarrow{x} p)$.
 	\end{enumerate} 	
 	The first claim follows from the fact that $R$ is the set of states participating in at least one accepting run over $w$.
 	For the second claim, if this were not the case, then the number of runs from $R$ to $R$ would increase unboundedly, 
 	contradicting that $\P$ is finitely ambiguous.
 	Thus for any state $p \in R$ there exists a unique run over $y$ from $p$ to some state in $R$, which is written $p'$.
 	To prove the third claim, pick a state $p \in R$ and note that
 	$
 	\P(q_{in} \xrightarrow{xy} p') = \P(q_{in} \xrightarrow{x} p) \cdot \P(p \xrightarrow{y} p') \le \P(q_{in} \xrightarrow{x} p)
 	$.
 	This reduces the analysis to two cases. On one hand, $p \le p'$ and then 
 	$
 	\P(q_{in} \xrightarrow{xy} p) \le \P(q_{in} \xrightarrow{xy} p') \le \P(q_{in} \xrightarrow{x} p)
 	$.
 	On the other hand, $p > p'$ and then there exists a state $q$ in $R$ such that $q \le p$ and $p \le q'$.
 	This is because for any state $r \in R$ there exists a unique run over $y$ from $r$ to some state in $R$.
  	It follows that	$\P(q_{in} \xrightarrow{xy} p) \le \P(q_{in} \xrightarrow{xy} q') \le \P(q_{in} \xrightarrow{x} q) \le \P(q_{in} \xrightarrow{x} p)$.
  	
 	The last claim implies the result, with the same calculations as for the proof of Lemma~\ref{lem:pumping_emptiness}.
 \end{proof}
 
 With the previous bounds in hand, we can study the computational
 complexity of the emptiness problem for various classes of finitely
 ambiguous probabilistic automata.  For each fixed positive integer
 $k$ we consider the class of $k$-ambiguous probabilistic automata.
 More generally, we can let the ambiguity of an automaton depend on
 the number $n$ of states: we consider for each fixed polynomial $P$
 the class of all automata that have ambiguity at most $P(n)$.  We
 call this the class of automata of $P$-finite ambiguity.  
 We emphasise that $P$-finite ambiguity is not the same as polynomial ambiguity:	
 for each $P$, the class of automata of $P$-finite ambiguity is a subclass of finitely ambiguous automata.
 (Recall that the ambiguity can be exponential in the number of states in general.)
 \smallskip
 
 \begin{theorem}\label{thm:complexity}\hfill
 	\begin{itemize}
        \item For each fixed positive integer $k$, the emptiness problem
          for the class of $k$-ambiguous probabilistic automata is in
          $\NP$.
        \item For each fixed polynomial $P$, the emptiness problem for
          the class of probabilistic automata with $P$-finite
          ambiguity is in $\PSPACE$.  This problem is $\PSPACE$-hard
          already in case $P(n)=n$.
        \item The emptiness problem for the class of finitely
          ambiguous probabilistic automata is in $\NEXPTIME$ and is
          $\PSPACE$-hard.
 	\end{itemize}
 \end{theorem}
 
 \begin{proof}
The algorithm for all three cases exploits Lemma~\ref{lem:pumping_emptiness} and Theorem~\ref{theo:pumping_emptiness} to guess and check a word witnessing that the threshold language is non-empty. 

For a $k$-ambiguous probabilistic automaton $\P$ we know by Lemma~\ref{lem:pumping_emptiness} 
that a witness for checking whether $L^{> c}(\P) \neq \emptyset$ is of polynomial length in $\P$ and, therefore, 
we can guess a word $w$ of appropriate length and check whether $\P(w) > c$ in polynomial time,
implying that the emptiness problem is in $\NP$.

Similarly, for finitely ambiguous $\P$ we know by Theorem~\ref{theo:pumping_emptiness} that the witness is of length at most exponential, 
so we can guess $w$ and check whether $\P(w) > c$ in $\NEXPTIME$.
 	
To show that emptiness is in $\PSPACE$ for probabilistic automata of $P$-finite ambiguity, 
one can guess a word $w$ ``on the fly'' of exponential length and check whether $\P(w) > c$.
The problem here is that the value $\P(w)$ (written in binary) could be of size exponential in the number of states of $\P$. 
To check whether $\P(w) > c$ with polynomial space one can guess $w$, and keep a set of counters $\{c_t^i\}$ that stores 
how many times each transition $t$ is used on the $i$-th run of $\P$ over $w$. 
Since $w$ is of length at most exponential and $\P$ has at most $P(n)$ accepting runs, then we need polynomially many counters, 
each with at most polynomially many bits, namely, polynomial space to store these counters during the simulation of $\P$ over $w$. 
After we conclude guessing $w$, we can construct a polynomial-size circuit that receives $\{c_i^t\}$ and outputs $\P(w)$.
Checking whether the value of the circuit is greater or equal than a constant $c$ can be solved in $\PSPACE$,
since both addition and multiplication are in $\NC$ and hence can be done in polylogarithmic space.
 	
Next we consider a fixed polynomial $p(n)=n$, and prove $\PSPACE$-hardness of emptiness for the class of probabilistic automata 
of $p(n)$-bounded ambiguity.
The proof is by reduction from the emptiness problem of the intersection of a finite collection of deterministic finite automata: 
given as input a collection of deterministic finite automata, does there exist a word accepted by each of them? 
This problem has been shown $\PSPACE$-complete in~\cite{K77}.
Given $N$ deterministic automata, we construct a probabilistic automaton $\P$ 
containing a copy of each deterministic automata and with a distribution of initial states 
assigning probability $\frac{1}{N}$ to the initial state of each automaton.
The probabilistic automaton $\P$ is $N$-ambiguous (note that $N$ is at most the number of states of $\P$), 
and there exists a word $w$ such that $\P(w) = 1$ if and only if there exists a word accepted by each of the $N$ deterministic automata.
\end{proof}
 
The aim of the last section is to give better algorithms for the $k$-ambiguous case: 
in particular, we show that the emptiness problem is in quasi polynomial time for $2$-ambiguous probabilistic automata.

%% file: algorithms.tex
This section is devoted to the construction of algorithms for both the emptiness problem 
and approximating the value of finitely ambiguous probabilistic automata.
The first step is a reduction to a multi-objective optimisation problem that we call the stochastic path problem.
We construct algorithms for this problem, relying on recent progress in the literature on multi-objective optimisation problems,
and thus obtain algorithms for finitely ambiguous probabilistic automata.

\subsection{The Stochastic Path Problem}
\label{subsec:stochastic_path_problem}

The stochastic path problem is an optimisation problem on
multi-weighted graphs.
It is parametrised by a positive integer constant $k$,
giving rise to the $k$-stochastic path problem.
An instance is a triple consisting of an
acyclic $k$-weighted graph $G$ and two vertices $s$ and $t$.  A
$k$-weighted graph is given by a set of vertices~$V$ of size $n$ and a
set of weighted edges $E \subseteq V \times (\Q \cap [0,1])^k \times V$. 
Note that the same pair of vertices $(v,v')$ can have several edges between them and 
the weight of an edge is a $k$-tuple of rational numbers between $0$ and $1$. 

A path $\pi$ in $G$ is a sequence of
consecutive edges, and the set of feasible solutions of the problem
are all paths from $s$ to $t$.
We let $(p_1(\pi),\ldots,p_k(\pi))$ denote the componentwise product of the weight vectors along the edges of~$\pi$.
In other words, the weight of a path on component $i$ is the product of the weights of each edge along $\pi$ on component $i$.
In our applications we think of each component of a weight vector of
an edge as the probability of a single event, and each component of a
weight vector of a path as the probability of a sequence of events.
The value of the path $\pi$, written $\val(\pi)$, is obtained by
summing each component of the weight vector of the path: $\val(\pi) = \sum_{i=1}^k p_i(\pi)$.

\begin{figure}
	\centering
	\begin{minipage}{.4\textwidth}
		\centering
		\includegraphics[width=\linewidth]{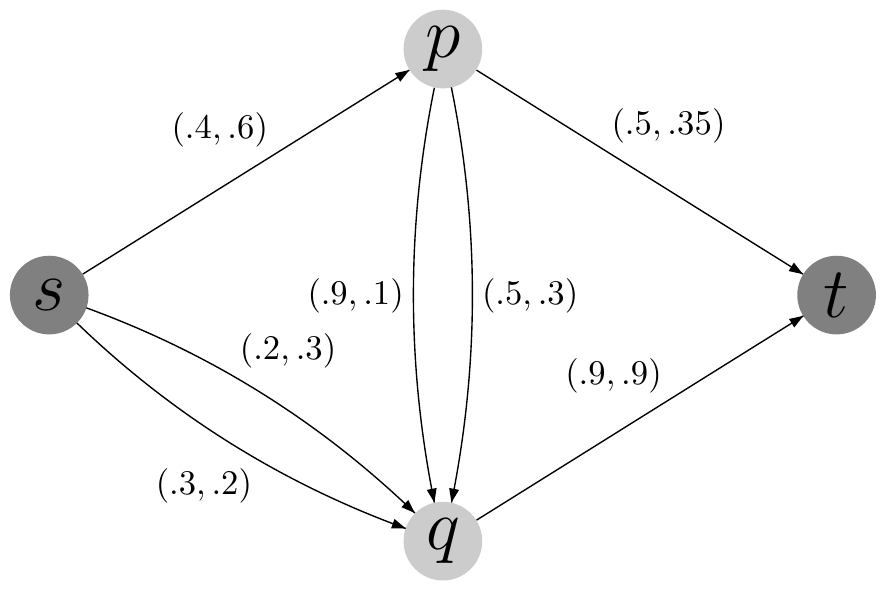}
	\end{minipage} \;\; 
	\begin{minipage}{.4\textwidth}
		\centering
		\includegraphics[width=.8\linewidth]{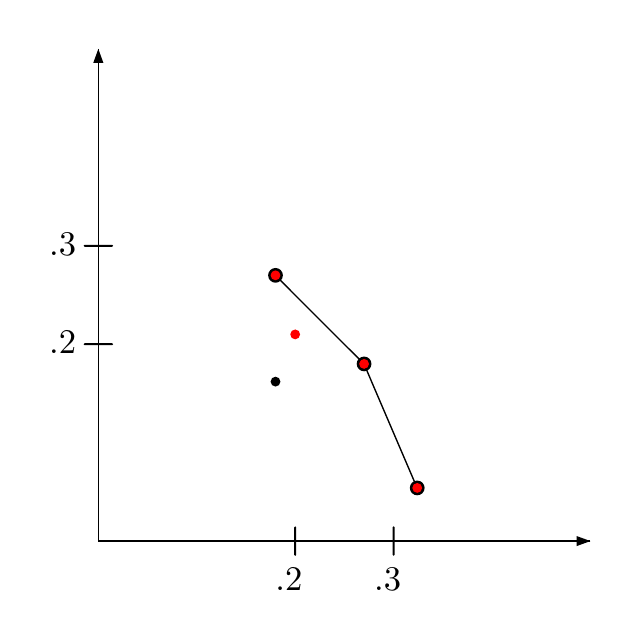}
	\end{minipage}
	\caption{\label{fig:example}An instance of the $2$-stochastic path problem on the left, and the values of all paths from $s$ to $t$ on the right.
		The four red dots are the Pareto curve, and the three connected red
		dots the convex Pareto curve.}
\end{figure}

As a running example, on the left-hand side of Figure~\ref{fig:example} we represent an instance of the $2$-stochastic path problem.  There are five paths from
$s$ to $t$, and their values are plotted in the right-hand side.  For
instance, the path $s,p,q,t$ using the left edge from $p$ to $q$ has
weight $(.4 \times .9 \times .9,\ .6 \times .1 \times .9) = (.324,.054)$,
so its value is $.324 + .054 = .378$. 

The objective of the $k$-stochastic path problem is to find the path with maximal value.
The decision problem associated with the $k$-stochastic path problem is the following:
\begin{framed}
	\textbf{The $k$-stochastic path problem}: 
	given a $k$-weighted graph $G$, two vertices $s$ and $t$ and a
	threshold $c$ in $\Q \cap [0,1]$, does there exist a path $\pi$ from
	$s$ to $t$ in $G$ whose value is at least $c$, \textit{i.e.} such that
	$\val(\pi) > c$?
\end{framed}

Towards finding efficient algorithms and approximations of $k$-ambiguous probabilistic automata, we show a polynomial time reduction from the emptiness problem of $k$-ambiguous probabilistic automata to the $k$-stochastic path problem.
Intuitively, the reduction consists in constructing the powerset graph of the paths, restricting to at most $k$ paths.

\begin{lemma}\label{lem:reduction_graph}
Fix $k \ge 2$.
There exists a polynomial-time reduction from the emptiness problem of $k$-ambiguous probabilistic automata 
to the $k$-stochastic path problem.
Given a $k$-ambiguous probabilistic automaton $\P$, 
the reduction constructs an instance of the $k$-stochastic path problem $(G_\P, s, t)$ satisfying the two following properties.
\begin{enumerate}
	\item for any word $w$ there exists a path $\pi$ in $G$ from $s$ to $t$ such that 
	$\P(w) \le \val(\pi)$,
	\item for any path $\pi$ in $G$ from $s$ to $t$ there exists a word $w$ such that 
	$\val(\pi) \le \P(w)$.
\end{enumerate}
\end{lemma}

\begin{proof}
Let $\P = (Q, q_{in}, \Delta, F)$ be a $k$-ambiguous probabilistic automaton with $n$ states.
The set of vertices of the $k$-weighted graph $G_\P$ is defined as $Q^k \times \{0, \ldots, n^k\} \times \{0,1\}^{k\times k}$ where $\{0,1\}^{k\times k}$ is the set of $k \times k$ matrices over $\{0,1\}$, plus a special source vertex $s$ and a special target vertex $t$.

Intuitively, being in the vertex $((q_1,\ldots,q_k), \ell, M)$ means that we are simulating $k$ runs which are now in the states $(q_1,\ldots,q_k)$,
that the run so far has length $\ell$, and the matrix $M$ indicates which pairs of runs are different: 
$M(i, j) = 1$ if and only if the $i$-th run is different from the $j$-th run.

We define the set of edges of $G_\P$.
\begin{itemize}
	\item For the source vertex, there is an edge from $s$ to $((q_{in},\ldots,q_{in}),0,\mathbf{0})$ with weight $(1,\ldots,1)$, 
	where $\mathbf{0}$ is the zero matrix.
	\item There is an edge from $((q_1,\ldots,q_k), \ell, M)$ to $((q'_1,\ldots,q'_k), \ell+1, M')$ with weight $(p_1,\ldots,p_k)$ 
if there exists a letter $a$ such that for each $i \in [1,k]$ we have $\Delta(q_i,a)(q'_i) = p_i$,
and $M'(i,j) = 1$ if and only if $M(i,j) = 1$ or $q_i' \neq q_j'$.
	\item There is an edge from $((q_1,\ldots,q_k), \ell, M)$ to $t$ with weight $(p_1,\ldots,p_k)$ 
where for each $i \in [1,k]$ we have $p_i = 1$ if $q_i \in F$ and $M(i, j) = 1$ for every $j < i$, and $p_i = 0$ otherwise.
\end{itemize}
Note that $G_\P$ is acyclic and of size polynomial in $\P$ given that $k$ is fixed.

We prove the correctness of the construction.
Let $w$ be a word. Thanks to Lemma~\ref{lem:pumping_emptiness}, 
we can assume without loss of generality that $w$ has length at most $n^k$.
Its set of accepting runs induces a path $\pi$ in $G_\P$ from $s$ to $t$ with $\val(\pi) = \P(w)$.
Conversely, a path $\pi$ in $G_\P$ from $s$ to $t$ corresponds to a set of accepting runs for some word $w$ with $\val(\pi) \leq \P(w)$.
\end{proof}

\subsection{Approximating the Value in Polynomial Time}
\label{subsec:fptas}

Multi-objective optimisation problems have long been studied; see~\cite{PY00} and~\cite{DY08} among many others.  
Since there is typically no single best solution, a natural notion for multi-objective
optimisation problems is \textit{Pareto curves}, which comprise sets 
of \textit{dominating} solutions.
To make things concrete, we illustrate the notion of Pareto curves on
the $k$-stochastic path problem.  We fix an instance $(G,s,t)$ of the
$k$-stochastic path problem.  A Pareto curve is a set of paths $\P$
such that for every path $\pi$, there exists a path $\pi'$ in $\P$
dominating~$\pi$, \textit{i.e.} such that for all $i$ in
$[1,k]$, we have $p_i(\pi) \le p_i(\pi')$.
In Figure~\ref{fig:example}, we can see that the Pareto curve of our running example is given by the four red dots.  
In dimension $2$ dominating means being to the right and higher, so only one path (represented by the black dot) is dominated by others.
Unfortunately, the size of Pareto curves in discrete multi-objective
optimisation problems is exponential in the worst case, 
motivating two relaxations: convex and approximate Pareto curves.

\vskip1em
A \textit{convex Pareto curve} is a set of paths $\C$ such that for every path $\pi$,
there exists a family of paths $\pi_1,\ldots,\pi_m \in \C$ such that
$\pi$ is dominated by a convex combination of $\pi_1,\ldots,\pi_m$ in
the sense that there exist non-negative coefficients
$\lambda_1,\ldots,\lambda_m$ that sum to~$1$ such that $p_i(\pi) \leq \sum_j \lambda_j p_i(\pi_j)$ for all components $i$ in
$[1,k]$.

Convex Pareto curves have been studied in a general setting by~\cite{DY08}. 
They are in general smaller than Pareto curves, yielding efficient algorithms for convex optimisation problems.  

In Figure~\ref{fig:example}, there exists a convex Pareto curve
consisting of only three paths, the fourth one being dominated a
convex combination of two other paths.  The figure connects the three
dots, showing what is called the Pareto front.

\vskip1em
Fix $\varepsilon > 0$,
an \textit{$\varepsilon$-Pareto curve} is a set of paths $\C$ such that for
every path $\pi$, there exists a path $\pi'$ in $\C$ such that for all
$i$ in $[1,k]$, we have $p_i(\pi) \le (1
+ \varepsilon) \cdot p_i(\pi')$.

The notion of approximate Pareto curves is very appealing for two reasons: first, knowing an approximate Pareto curve usually gives an
approximately optimal solution, and second, a very general result of~\cite{PY00} shows that in most
multi-objective optimisation problems, there exists a polynomially
succinct approximate Pareto curve.

\vskip1em
The two relaxations can be combined: 
an \textit{$\varepsilon$-convex Pareto curve} is a set of paths $\C$ such that for every path $\pi$,
there exists a family of paths $\pi_1,\ldots,\pi_m \in \C$ and non-negative coefficients $\lambda_1,\ldots,\lambda_m$ that sum to~$1$ such that 
$p_i(\pi) \leq (1 + \varepsilon) \sum_j \lambda_j p_i(\pi_j)$ for all components $i$ in $[1,k]$.

\vskip1em
The following result shows how to find a $(1 + \varepsilon)$-approximation of the value of a $k$-ambiguous probabilistic automaton $\P$.
\begin{theorem}\label{thm:fptas}
	There exists an algorithm which given an instance of the $k$-stochastic path problem and $\varepsilon > 0$,
	returns an $\varepsilon$-convex Pareto curve in time polynomial in the instance and~$\frac{1}{\varepsilon}$.
\end{theorem}
\begin{proof}
We rely on general results of~\cite{PY00}, 
which give a sufficient condition for the existence of a polynomial time algorithm constructing 
an $\varepsilon$-convex Pareto curve in time polynomial in the instance and~$\frac{1}{\varepsilon}$:
it is enough to construct an algorithm solving the exact version in pseudo-polynomial time.
Recall here that an algorithm is pseudo-polynomial if it runs in polynomial time when the numerical inputs are given in unary.

In our case, the exact $k$-stochastic path problem reads:
given an instance $(G,s,t)$ and a value $c$ in $[0,1] \cap \Q$, does there exist a path $\pi$ in $G$ from $s$ to $t$ 
such that $\val(\pi) = c$?
Let $n$ be the number of vertices of $G$.
If all transition probabilities have size $B$ (in unary),
then it is enough to consider paths such that each weight has size $n \cdot B$ (in unary).
Hence one can fill in a polynomially large table indexed by $(p,q,p_1,\ldots,p_k)$, which checks
for the existence of a path from $p$ to $q$ of weights $(p_1,\ldots,p_k)$ of size $n \cdot B$ (in unary).
\end{proof}

The algorithm of Theorem~\ref{thm:fptas} for the $k$-stochastic path problem yields a polynomial time algorithm to 
approximate the value of a $k$-ambiguous probabilistic automaton. 

\begin{theorem}\label{thm:fptas_pa}
	There exists an algorithm which given a $k$-ambiguous probabilistic automaton and $\varepsilon > 0$, 
	returns a $(1 + \varepsilon)$-approximation of the value in time polynomial in the size of the automaton and~$\frac{1}{\varepsilon}$,
	and more specifically a value $\Output$ such that
	\[
	\Output \le \val(\P) \le (1 + \varepsilon) \cdot \Output.
	\]	
\end{theorem}

\begin{proof}
	Given a $k$-ambiguous probabilistic automaton $\P$, the algorithm for finding a $(1 + \varepsilon)$-approximation of $\val(\P)$ is as follows:
	\begin{enumerate}
		\item construct the instance $(G_\P,s,t)$ of the $k$-stochastic path problem using Lemma~\ref{lem:reduction_graph}.
		\item construct an $\varepsilon$-convex Pareto curve $\C$ for $(G_\P,s,t)$ thanks to Theorem~\ref{thm:fptas}.
		\item return $\Output := \max_{\pi \in \C} \sum_{i \in [1,k]} p_i(\pi)$.
	\end{enumerate}
	The first inequality is a direct consequence of Lemma~\ref{lem:reduction_graph} given that for every path $\pi$ in $G_\P$, 
	there exists a word $w$ such that $\val(\pi) \leq \P(w)$, so $\Output \leq \val(\P)$.

	For the second inequality, consider a word $w$ that achieves $\P(w) = \val(\P)$. 
	By Lemma~\ref{lem:reduction_graph}, there exists a path $\pi$ such that $\P(w) \leq \val(\pi)$. 
	Since $\C$ is an $\varepsilon$-Pareto curve, there exists a path $\pi' \in \C$ such that for all $i$ in $[1,k]$,
	we have $p_i(\pi) \le (1 + \varepsilon) \cdot p_i(\pi')$.
	It follows that $\P(w) \le (1 + \varepsilon) \cdot \val(\pi') \le (1 + \varepsilon) \cdot \Output$.
\end{proof}

It is interesting to compare the positive result of Theorem~\ref{thm:fptas_pa} to the negative result of Theorem~\ref{thm:hardness_approximation}.
The key difference is in fixing the ambiguity, which allows us to go from intractable to tractable.

\subsection{A Quasi-Polynomial Time Algorithm for 2-ambiguous Probabilistic Automata}

The previous result show that one can $(1 + \varepsilon)$-approximate the value of $k$-ambiguous probabilistic automaton in polynomial time.
This is however not enough to decide the emptiness problem.
In this direction, Theorem~\ref{thm:complexity} shows that for any fixed $k$ 
the emptiness problem of $k$-ambiguous probabilistic automata is in $\NP$.
We show that for $k = 2$ there exists a quasi-polynomial time algorithm for the emptiness problem. 
For this, we start by constructing a quasi-polynomial time algorithm for the $2$-stochastic path problem.

\begin{theorem}\label{thm:quasi_poly}
	There exists an algorithm which given an instance of the $2$-stochastic path problem, returns a convex Pareto curve in quasi-polynomial time.
\end{theorem}

The benefit of fixing $k = 2$ lies in the existence of a quasi-polynomial bound on the size of convex Pareto curves.
More precisely, if $(G,s,t)$ is an instance of the $2$-stochastic path problem with $n$ vertices,
then it can be shown that there exists a convex Pareto curve of size at most $n^{\log(n)}$.
This result was proved in ~\cite{G80}, and a matching lower bound was developed by~\cite{C83}.
Note that they use a different framework, called parametric optimisation: 
in the parametric shortest path problem each edge has cost $c + \lambda d$, where $\lambda$ is a parameter.
The length of the shortest path is a piecewise linear concave function of $\lambda$, 
whose pieces correspond to the vertices of the convex Pareto curve for the shortest path problem with weights $(c,d)$.
It is then easy to obtain an upper bound on the size of convex Pareto curves for the $2$-stochastic path problem by reducing it
to the parametric shortest path problem, mapping the weights $(p,q)$ to $(-\log(p),-\log(q))$.
Finally, the upper bound on the size of convex Pareto curves yields a quasi-polynomial time algorithm, by constructing them in a standard divide-and-conquer manner.

The algorithm of Theorem~\ref{thm:quasi_poly} yields a quasi-polynomial time algorithm for the emptiness problem of $2$-ambiguous probabilistic automata.

\begin{theorem}
	There exists a quasi-polynomial time algorithm for the emptiness problem of $2$-ambiguous probabilistic automata.
\end{theorem}
\begin{proof}
	Given a $2$-ambiguous probabilistic automaton $\P$ and a threshold $c$, an algorithm for deciding the emptiness of $\P$ is as follows:
	\begin{itemize}
		\item construct the instance $(G_\P,s,t)$ of the $2$-stochastic path problem using Lemma~\ref{lem:reduction_graph}.
		\item construct a convex Pareto curve $\C$ for $(G_\P,s,t)$ thanks to Theorem~\ref{thm:quasi_poly}.
		\item check whether $\Output := \max_{\pi \in \C} \sum_i p_i(\pi) > c$.
	\end{itemize}
	To show the correctness of this algorithm, we first prove that $\P(w) \le \Output$ for every word $w$.
	Let $w$ be a word, thanks to Lemma~\ref{lem:reduction_graph}, there exists a path $\pi$ such that $\P(w) \le p_1(\pi) + p_2(\pi)$.
	Since $\C$ is a convex Pareto curve, there exists a convex combination of paths $\pi' = \lambda_1' \pi_1' + \lambda_2' \pi_2'$ in $\C$ such that 
	$p_1(\pi) \le p_1(\pi')$ and $p_2(\pi) \le p_2(\pi')$.
	Now, consider all convex combinations of paths in $\C$; by convexity of the sum function, 
	the maximum over this set is reached on some path $\pi_m''$, 
	so $p_1(\pi') + p_2(\pi') \le p_1(\pi_m'') + p_i(\pi_m'')$. 
	It follows that $\P(w) \le p_1(\pi_m'') + p_i(\pi_m'') \le \Output$.
	
	To conclude the proof of correctness, we show that $\Output \le \P(w)$ for some word~$w$. 
	Indeed, if $\pi$ is a path such that $\Output =  p_1(\pi) + p_2(\pi)$, then thanks to Lemma~\ref{lem:reduction_graph}
	there exists a word $w$ such that $p_1(\pi) + p_2(\pi) \le \P(w)$.
\end{proof}